\documentclass[twocolumn, trackchanges]{aastex631}
\usepackage{amsmath}

\shorttitle{Implications of `Oumuamua on Panspermia}
\shortauthors{Cao et al.}
\graphicspath{{./}{figures/}}

\begin{document}
\title{The Implications of `Oumuamua on Panspermia}

\author{David Cao}
\affiliation{Thomas Jefferson High School for Science and Technology \\
6560 Braddock Rd \\
Alexandria, VA 22312, USA}

\author[0000-0002-8864-1667]{Peter Plavchan}
\affiliation{Department of Physics and Astronomy, George Mason University \\
4400 University Dr. \\
Fairfax, VA 22030, USA}

\author{Michael Summers}
\affiliation{Department of Physics and Astronomy, George Mason University \\
4400 University Dr. \\
Fairfax, VA 22030, USA}

\begin{abstract}
Panspermia is the hypothesis that life originated on Earth from the bombardment of foreign interstellar ejecta harboring polyextremophile microorganisms. Since the 2017 discovery of the interstellar body ‘Oumuamua (1I/2017 U1) by the Pans-STARRS telescope, various studies have re-examined panspermia based on updated number density models that accommodate for `Oumuamua's properties. By utilizing ‘Oumuamua's properties as an anchor, we estimate the mass and number density of ejecta in the ISM ($\rho_{m}$\,[kg$\cdot$au$^{-3}]$ and $\rho_{n}$\,[au$^{-3}]$). We build upon prior work by first accounting for the minimum ejecta size to shield microbes from supernova radiation. Second, we estimate the total number of impact events $C_{n}$ on Earth after its formation and prior to the emergence of life ($\approx$ $0.8$\,Gyr). We derive a conditional probability relation for the likelihood of panspermia for Earth specifically of $<10^{-5}$, given a number of factors including $f_{B}$, the fraction of ejecta harboring extremophiles and other factors that are poorly constrained. However, we find that panspermia is a plausible potential life-seeding mechanism for (optimistically) {potentially} up to $\sim10^5$ of the $\sim10^9$ Earth-sized habitable zone worlds in our Galaxy. 

\end{abstract}

\section{Introduction} \label{sec:intro}

The recently discovered `Oumuamua merits a re-examination for the possibility of panspermia, the hypothesis that life seeded on Earth from the bombardment of life-bearing interstellar ejecta, and that life can be transferred from one celestial body to another. `Oumuamua’s hyperbolic orbit ($e \approx 1.2$) and extreme solar system inclination angle ($\theta \approx 123^\circ$) demonstrated that the body originated from the ISM, the first large interstellar object to ever be observed \citep{Jewitt_2017, Taylor1996}. This implies that `Oumuamua ejected from a planetary system and happened to pass by Earth, a prime example of what could have been a mechanism for panspermia \citep{Trilling_2017}.

However, panspermia is an extraordinarily difficult theory to quantitatively model and assess \citep{TEPFER2008756, Mcnichol2012, Kawaguchi2019, Gobat_2021, Grimaldi_2021}. It is a complex, multifaceted concept requiring both physical and biological parameters that are empirically poorly constrained. The mechanism for panspermia requires an abundance of rocky ejecta from other planetary systems that reach the ISM and later bombard the Earth. In addition, said ejecta must originate from a life-bearing body, and the extremophiles living on the ejecta must endure harsh conditions including but not limited to high accelerations and supernova gamma radiation. 

Life-bearing ejecta may collide with objects (including the Earth) during the process of panspermia. This subjects extremophiles to high accelerations. Previously, two lab experiments tested the resistance of B. Subtilis, an extremophile, to centrifuges and lead-pellet rifles. Forces experienced by microbes during the experiment were 2.5 to 5 times greater than simulated collisions in space, and the extremophiles survived \citep{MASTRAPA20011}. Hence, intermediate collisions do not hinder the prospect of panspermia.

In contrast, gamma ray radiation from nearby supernovas are more threatening to the survival of extremophiles in the ISM. Natural protection such as carbonated skin and rocky pores are insufficient to prevent the penetration of gamma rays - extremophiles must be shielded by a certain depth from the ejecta itself \citep{10.1111/j.1365-2966.2004.07287.x}. Various studies confirm the potentially harmful, deleterious impact of intense radiation on extremophiles such as \citet{Horneck2001}. Other studies have also expressed reasonable concern, such as \citet{doi:10.1126/science.160.3829.754, 10.1111/j.1365-2966.2004.07287.x, Brunton_2023, Thomas_2023}. 

In light of the 2017 `Oumuamua flyby, models from several studies imply higher ISM ejecta flux rates. For example, \citet{Levine_2021, doi:10.1146/annurev-astro-071221-054221, seligman2023interstellar} approximate the number density of `Oumuamua-sized ejecta in the ISM as $\sim 10^{-1}$\,au$^{-3}$, and \citet{Do_2018} derives an even greater number density of `Oumuamua-like ejecta in the ISM $\approx 0.2$\,au$^{-3}$. Additionally, \citet{Do_2018} suggests that $\sim 10^{15}$ objects are present in the Oort Cloud alone, which actively produces ISM ejecta. These findings imply that the flux rate of ISM ejecta is orders of magnitude higher than previously thought \citep{Meech2017}. Additionally, given `Oumuamua's size, \citet{siraj20222019} estimates the size distribution of ISM ejecta as $N(>R) \sim kR^{1-q}$ using Poisson distributions, where the distribution slope $q \approx 3.6 \pm 0.5$. This implies a significantly higher population of larger asteroids because $q < 4$, for $q = 4$, the scale-free power law exponent, all ejecta sizes contribute the same mass in a given volume of space. This is potentially significant because ejecta of larger sizes (at least in the meter range) are necessary to sustain microbial life during panspermia \citep{Ginsburg_2018}. 

The expected increase in the flux rate of ISM ejecta following the discovery of `Oumuamua warrants a closer examination for the potential for panspermia. This is because a higher ISM flux rate implies more interactions between bodies through bombardments of potentially life-seeding ejecta \citep{Ginsburg_2018, Moro-Martín_2022}. Recently, several studies have directly explored the implications of `Oumuamua on panspmeria. Based on the model from \citet{Moro-Martín_2022}, the total number of interstellar ejecta captured by a planetary system exceeds $10^{11}$ over its lifetime. \citet{Ginsburg_2018} proposed an estimate for the capture rate of ISM ejecta given constraints such as the anchor ISM ejecta number density (based on `Oumuamua) and ejecta size (also based on `Oumuamua). Both values are given by `Oumuamua's $\approx 100$\,m size and $\sim 10^{-1}$\,au$^{-3}$ ISM ejecta number density. Within a $\sim$ 1\,Gyr time frame, \citet{Ginsburg_2018} estimates that over $10^{6}$ Enceladus-sized interstellar objects with size $\sim 100$\,km have been captured by our galaxy. This implies a greater number of smaller ejecta that have also been captured \citep{siraj20222019}. Though the Milky Way is spacious therefore the chance that solar systems and planets capture ejecta seem insignificant, \citet{Ginsburg_2018} proposed that bombardment rates are sufficiently large so that the entire Milky Way could be indirectly exchanging biotic life. In addition, \citet{Grishin_2019} showed that $\sim$ $10^{4}$ `Oumuamua-like ejecta were likely captured by the Solar System amidst planetary formation. In a cluster environment such as a protoplanetary disk, \citet{Grishin_2019} finds that the rate of capture of $\sim$ 1\,km objects is 50 times higher than previous estimates. This subsequently increases the capture rate of smaller ejecta by several magnitudes and therefore enhances the viability of panspermia. 

The subsequent discovery of Comet 2I/Borisov in 2019 may also provide useful context about panspermia. Its larger size of between 0.2\,km and 0.5\,km \citep{Jewitt_2020} would better protect microorganisms from gamma ray radiation, and the dust trails left behind by the comet could serve as a non-collisional mechanism for material transfer (${\S}$\ref{subsec: 4.5}). We do not incorporate 2I/Borisov into our main study due to a lack of existing estimates of occurrence rates in this size range (${\S}$\ref{sec2:Borisov}).

In this work, we build on previous studies by assessing both physical and biological constraints on panspermia, in particular the total number of impact events on Earth and minimum radiation shielding depth. We begin by developing a model for the ISM ejecta number density and mass density in Section \ref{sec:densities}. Since `Oumuamua is the first object observed to originate from the ISM, we use its size as an anchor to model both the number density and mass density of ISM ejecta. We approximate the mass flux and flux density of ejecta in the ISM and derive the total number of collisions on Earth prior to the fossilized presence of life. In Section \ref{sec:shielding}, we find the minimum ejecta size needed to protect extremophiles from supernova radiation in the ISM and calculate the total number of ejecta captured given that restriction. In Section \ref{sec:summary}, we place our results in the context of panspermia, compare our results with other studies, address remaining uncertainties, present analysis of the findings from the Vera Rubin Observatory as an area of future work, and discuss the plausibility for the transfer of life via dust-grained sized ejecta from an interstellar object during a flyby. In Section \ref{sec:concf}, we summarize our findings and present possible avenues for future work. 

\section{Total Impact Events \& Collision Mass} \label{sec:densities}

In Section \ref{subsec:2.1}, we model the number density and flux rate of ejecta in the ISM to achieve the total number of impact events during the time period between Earth’s formation and the earliest fossilized evidence for the emergence of life, $\approx$ 0.8\,Gyr \citep{doi:10.1073/pnas.1517557112}. In Section \ref{subsec:2.3}, we model the mass density and mass flux of ejecta in the ISM to achieve the total collision mass within the same $\approx$ 0.8\,Gyr time interval. We present our results for the total impact events and total collision mass in Section \ref{sec2:results}. Both the total number of collisions and the total collision mass are essential metrics to evaluate since they directly impact the plausibility of panspermia; a relatively low result could indicate that panspermia is unlikely, whereas a high result indicates that panspermia is more probable. 

\subsection{Total Impact Events} \label{subsec:2.1}

We first model the ISM ejecta number density to estimate the total number of impact events on Earth prior to the earliest fossilized evidence for life. We model the ISM ejecta number density with respect to the size of ejecta by using `Oumuamua as an anchor. Ejecta number density and size are inversely proportional and can be written as

\begin{equation} \label{Eq:1}
N(r) = N_{o}\left(\frac{R_{o}}{r}\right)^{q},
\end{equation}

where $N(r)$\,[au$^{-3}]$ is the ISM ejecta number density of ejecta with size $r$\,[m], $N_{o}$ is the number density of `Oumuamua-like objects in interstellar space, $R_{o} \approx 100$\,m is the size of `Oumuamua, and $q$ is the model's distribution slope where $3.1 < q < 4.1$ \citep{Do_2018,  Bannister2019, siraj20222019}. We take $N_{o} \approx 0.1$\,au$^{-3}$ as a representative, order-of-magnitude value estimated from prior works:  \citet{Do_2018} estimated $N_{o} \approx 0.2$\,au$^{-3}$, the `Oumuamua ISSI Team estimated $\approx 0.1$\,au$^{-3}$, and \citet{Portegies} estimated a range of 0.004--0.24\,au$^{-3}$ with an expected value of 0.08\,au$^{-3}$. Throughout this study, for the number density power-law exponent, we are primarily interested in $q = 3.6$ since it is the numerical average of the estimated range, but we retain it as a free model parameter. 

For $q > 0$, integrating $|\frac{dN(r)}{dr}|dr$ gives the total number of ejecta in 1\,au of space. For now, we assume an upper bound of $r = 1.5R_{\oplus}$. We do this because the number density model in Equation (\ref{Eq:1}) is meaningless at the extremities, since terrestrial objects have radii $< 1.5R_{\oplus}$ \citep{Rogers_2015}.

Next, given upper bound ejecta size $r = 1.5R_{\oplus}$, we have

\begin{equation} \label{Eq:2}
\rho_{n}(x) = \int_{x}^{1.5R_{\oplus}} {\left|\frac{dN(r)}{dr}\right|}dr,
\end{equation}

which gives
 
\begin{equation} \label{Eq:3}
\rho_{n}(x) = N(x) - N(1.5R_{\oplus}),
\end{equation}

where $\rho_{n}(x)$\,[au$^{-3}]$ is the ISM ejecta number density (note, the sign flip from the absolute value). Next, to find the number of impact events, we first evaluate the ejecta flux in the ISM,

\begin{equation} \label{Eq:4}
\Phi_{n}(\rho_{n}(x)) = \rho_{n}(x)v_{o}\hat{t},
\end{equation}

where $v_{o}$\,[m$\cdot$s$^{-1}]$ is the velocity of the ejecta based on `Oumuamua ($\approx$ 26\,km$\cdot$s$^{-1}$ relative to the local standard of rest) and $\hat{t}$\,[s$\cdot$yr$^{-1}]$ is the unit conversion constant for time where $\hat{t} \approx 3.2 \times 10^{7}$\,s$\cdot$yr$^{-1}$ to re-express the velocity in km\,yr$^{-1}$ \citep{Do_2018}. The total number of impact events is therefore equivalent to

\begin{equation} \label{Eq:5}
C_{n}(\Phi_{n}) = \Phi_{n}\tau\sigma_{\oplus}h_{\oplus}h_{\odot},
\end{equation}

where $\tau$\,[Gyrs] is the time between Earth's formation and the earliest fossilized evidence for life, $\sigma_{\oplus}$\,[m$^{2}]$ is the cross-sectional area of the Earth, $h_{\oplus}$ is the gravitational focusing value of the Earth, and $h_{\odot}$ is the gravitational focusing value of the Sun. We take $\tau \approx 0.8$\,Gyrs as a representative order-of-magnitude estimate based upon \citet{Ohtomo}, which details evidence for biogenic graphite in metasedimentary rock 3.7 Gyr old following the formation of Earth $\approx$ 4.5\,Gyr ago, though an exact consensus among different methodologies is yet to be reached \citep{Pearce2018-dv}. We take $\sigma_{\oplus} = \pi R_{\oplus}^{2} \approx 1.3 \times 10^{14}\,$m$^{2}$, $h_{\oplus}=1 + {v_{\oplus}^2}/{v_{i}^2} \approx$ $1.18$ where the escape velocity of Earth $v_{\oplus}= \sqrt{2GM_{\oplus}/R_{\oplus}}$\,km$\cdot$s$^{-1}$ $\approx$ 11\,km$\cdot$s$^{-1}$, and $h_{\odot} = 1 + {v_{\odot}^2}/{v_{i}^2}\approx 3.63$ where the escape velocity of the Sun $v_{\odot} = \sqrt{2GM_{\odot}/1.5\times 10^{8}\,km}\,$km$\cdot$s$^{-1}$$ \approx 42\,$km$\cdot$s$^{-1}$. Note $1.5\times10^{8}\,$km $\approx 1\,$au. \citep{10.1111/j.1365-2966.2006.11349.x}. 

\begin{figure}[ht!]
\includegraphics[scale=0.4]{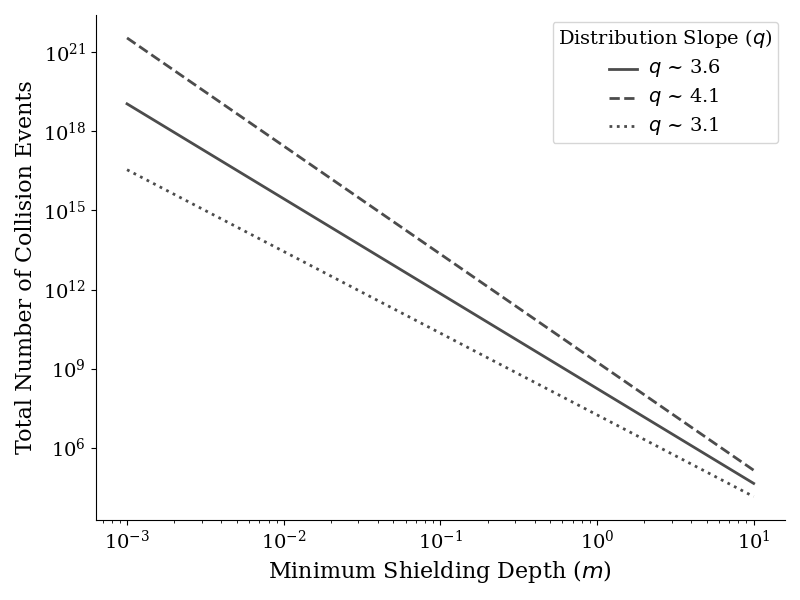} 
\caption{Total number of impact events $C_{n}(\Phi_{n})$ from the time of the formation of the Earth to the earliest fossilized evidence for life ($\approx$ $0.8$\,Gyrs) as a function of the minimum ejecta shielding depth $x$\,[m] with a range of $10^{-3} < x < 10$ m \citep{Ohtomo}.\label{fig:f1}}
\end{figure} 

\begin{deluxetable*}{cccccc}
\label{ISMdensities}
\tablenum{1}
\tablecaption{Modeled ISM densities and ejecta collision statistics \label{tab:ism densities}}
\tablewidth{0pt}
\tablehead{
\colhead{Min. Shielding $x$} & \colhead{Density $\rho_{m}$ $r \le \infty$} & \colhead{Density $\rho_{m}$ $r \le 1.5R_{\oplus}$} & \colhead{$\%$ Difference} & \colhead{Total Coll. Mass $C_{m}$} &  \colhead{Total Coll. Events $C_{n}$} \\
\colhead{m} & \colhead{kg$\cdot$au$^{-3}$} & \colhead{kg$\cdot$au$^{-3}$} & \colhead{\%} & \colhead{kg} & \colhead{number}
}
\decimalcolnumbers
\startdata
0.001 & $1 \times 10^{12}$ & $1 \times 10^{12}$ & $3 \times 10^{-4}$ & $1 \times 10^{14}$ & $1 \times 10^{19}$ \\
0.01 & $3 \times 10^{11}$ & $3 \times 10^{11}$ & $1 \times 10^{-3}$ & $3 \times 10^{13}$ & $3 \times 10^{15}$ \\
0.1 & $6 \times 10^{10}$ & $6 \times 10^{10}$ & $4 \times 10^{-3}$ & $7 \times 10^{12}$ & $7 \times 10^{11}$ \\
1 & $2 \times 10^{10}$ & $2 \times 10^{10}$ & $2 \times 10^{-2}$ & $2 \times 10^{12}$ & $2 \times 10^{8}$ \\
10 & $4 \times 10^{9}$ & $4 \times 10^{9}$ & $6 \times 10^{-2}$ & $4 \times 10^{11}$ & $4 \times 10^{4}$ \\
\textbf{Water-ice (6.6)} & \textbf{$5 \times 10^{9}$} & \textbf{$5 \times 10^{9}$} & \textbf{$5 \times 10^{-2}$} & \textbf{$6 \times 10^{11}$} & \textbf{$2 \times 10^{5}$} \\
\enddata 
\tablecomments{Modeled data for the ISM mass density $\rho_{m}(x)$, total collision mass $C_{m}(\Phi_{m})$, and total number of impact events $C_{n}(\Phi_{n})$ from the time period of Earth's formation to the earliest fossilized evidence of life ($\approx$ $0.8$\,Gyrs) given $q = 3.6$ \citep{Ohtomo}. We estimate ISM mass densities for two ejecta size upper bounds: $r \le \infty$ and $r \le 1.5R_{\oplus}$. Results indicate that the bounds have negligible differences, hence the mass attributed by ejecta with $r \ge 1.5R_{\oplus}$\, is relatively small. Both $C_{m}(\Phi_{m})$ and $C_{n}(\Phi_{n})$ were calculated with $r \le 1.5R_{\oplus}$.}
\end{deluxetable*}
\vspace*{-\baselineskip}

\begin{figure}[ht]
\includegraphics[scale=0.4]{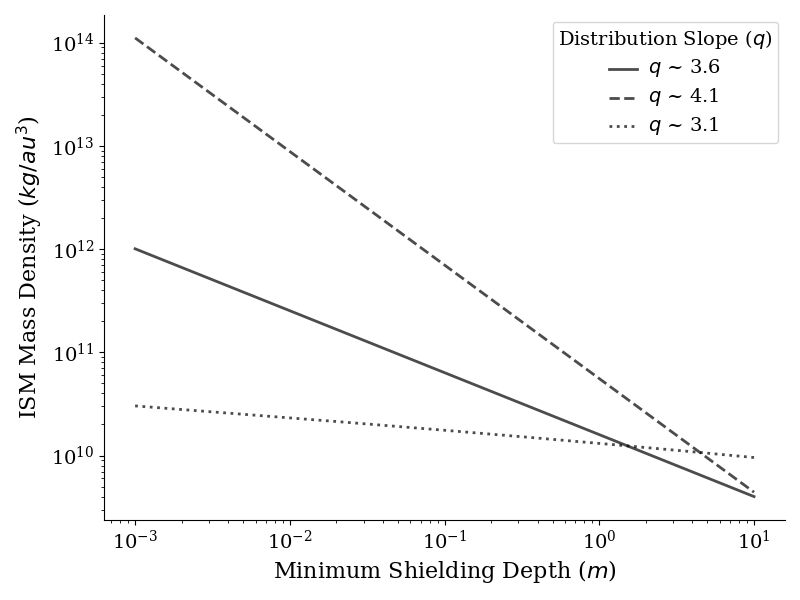}
\caption{Modeled ISM Mass Density $\rho_{m}(x)$\,[kg$\cdot$au$^{-3}]$ as a function of the minimum ejecta shielding depth $x$\,[m] with a range of $10^{-3} < x < 10$ m. \label{fig:f2}} 
\end{figure}

\subsection{Total Collision Mass} \label{subsec:2.3}

Next, we model the mass of ISM ejecta and apply our ejecta number density model in Section \ref{subsec:2.1} to determine the total collision mass that accumulated on Earth prior to the earliest fossilized evidence for life. 

We first model the mass of individual ejecta in comparison to that of `Oumuamua. Since ejecta mass is proportional to the cube of its size, we have

\begin{equation} \label{Eq:6}
M(r) = M_{o}\left(\frac{r}{R_{o}}\right)^{3},
\end{equation}

where $M(r)$\,[kg] is the mass of an ejecta with size $r$\,[m] and $M_{o}$\,[kg] is the mass of `Oumuamua up to $\approx 4 \times 10^{9}$\,kg given size $R_{o} \sim 100$\,m and an assumed order-of-magnitude density for `Oumuamua of $\rho_{o} \sim 1$\,g$\cdot$cm$^{-3}$ \citep{Do_2018, McNeill_2018}. 

Extremophiles, microorganisms that can endure harsh environments, are mandatory for life to survive transport in space for panspermia to be a feasible possibility. Extremophiles are susceptible to supernova gamma-ray radiation during panspermia. Therefore ejecta must have minimum radius $x$ to shield extremophiles buried in the ejecta beneath the surface at least $x$ m \citep{10.1111/j.1365-2966.2004.07287.x}. To find the total ISM mass density of ejecta with size range $x < r < 1.5R_{\oplus}$, we have

\begin{equation} \label{Eq:7}
\rho_{m}(x) = \int_{x}^{1.5R_{\oplus}} {M(r - x)\left|\frac{dN(r)}{dr}\right|}dr.
\end{equation}

Note that the expression in Equation (\ref{Eq:7}) is $M(r - x)$ and not $M(r)$ to accommodate for the aforementioned supernova gamma radiation that introduces a minimum ejecta depth where extremophiles can survive. After integration and simplification, we achieve the final equation

\begin{equation} \label{Eq:8}
\rho_{m}(x) = \left\{
\begin{array}{ll}
       qM_{o}N_{o}R^{q-3}_{o}(A - B), & q\neq0,1,2,3 \\ 
      \frac{1}{2}M_{o}N_{o}R_{o}^{-2}C& q=1 \\
       M_{o}N_{o}R_{o}^{-1}D & q=2 \\
       M_{o}N_{o}E & q=3 \\
\end{array} 
\right.
\end{equation}

where

\begin{equation} \label{Eq:9}
\begin{aligned}
 A = \frac{(1.5R_{\oplus})^{3-q}}{3-q} - \frac{3x(1.5R_{\oplus})^{2-q}}{2-q} \\ + \frac{3x^{2}(1.5R_{\oplus})^{1-q}}{1-q} + \frac{x^{3}(1.5R_{\oplus})^{-q}}{q},
\end{aligned}
\end{equation}

\begin{equation} \label{Eq:10}
 B = \left(x^{3-q}\right)\left(\frac{1}{3 - q} - \frac{3}{2 - q} + \frac{3}{1 - q} + \frac{1}{q}\right),
\end{equation}

\begin{equation} \label{Eq:10.1}
\begin{aligned}
 C =  \frac{2x^3}{1.5R_{\oplus}} +6x^2\ln(\frac{1.5R_{\oplus}}{x}) + 3x^2 \\ - 9R_{\oplus}x + (1.5R_{\oplus})^{2},
 \end{aligned}
\end{equation}

\begin{equation} \label{Eq:10.2}
\begin{aligned}
D = \frac{x^3}{(1.5R_{\oplus})^2}-\frac{6x^2}{1.5R_{\oplus}} + 6x\ln(\frac{x}{1.5R_{\oplus}})
\\ + 3x  + 3R_{\oplus},
\end{aligned}
\end{equation}

and

\begin{equation} \label{Eq:10.3}
\begin{aligned}
E = \frac{x^3}{(1.5R_{\oplus})^3} - \frac{9x^2}{4.5R_{\oplus}^2} + \frac{9x}{1.5R_{\oplus}} \\ + 3\ln(\frac{1.5R_{\oplus}}{x}) - \frac{11}{2}.
\end{aligned}
\end{equation}

Note the general form requires $q \neq 0, 1, 2, 3$ due to discontinuities in $A$ and $B$, but specific solutions for those $q$ values are provided.

Next, evaluating the ejecta mass flux in the ISM, we have

\begin{equation} \label{Eq:11}
\Phi_{m}(\rho_{m}(x)) = \rho_{m}(x)v_{o}\hat{t}.
\end{equation}

The total ejecta collision mass is therefore equivalent to

\begin{equation} \label{Eq:12}
C_{m}(\Phi_{m}) = \Phi_{m}\tau\sigma_{\oplus}h_{\oplus}h_{\odot}.
\end{equation}

\subsection{Results: Total Impact Events \& Collision Mass} \label{sec2:results}

Figure \ref{fig:f1} shows the effect of minimum ejecta size $x$ on the total number of impact events $C_{n}(\Phi_{n})$. As mentioned in Section \ref{sec:intro}, ejecta must be sufficiently large to protect extremophiles from supernova gamma radiation \citep{doi:10.1126/science.160.3829.754, Horneck2001, 10.1111/j.1365-2966.2004.07287.x, Brunton_2023, Thomas_2023}. Smaller $x$ leads to greater sensitivity to variations in our assumed model parameters $q$ and $N_o$ due to extrapolation for the number density of ejecta, $N(r)$. Conversely, increasing $x$ results in a decreased sensitivity to our choice of model parameters, as all values of $q$ for $C_{n}(\Phi_{n})$ converges.

Table \ref{tab:ism densities} lists specific values of the total number of impact events $C_{n}(\Phi_{n})$ given $q = 3.6$ and varying magnitudes of the minimum ejecta shielding depth $x$. We choose values for $x$\,[m], the minimum ejecta size, from $0.001$\,m to $10$\,m, least to most conservative. Any ejecta size smaller than $0.001$\,m is extremely unlikely to sustain life due to the harmful effects of UV radiation, and any size larger than $10$\,m can sustain interior survival conditions for extremophiles for on the order of $\sim$7.5\,Gyr in the ISM \citep[Eqn 2, ][]{Valtonen_2009}; see also \citet{Grimaldi_2021}. Since $C_{n}(\Phi_{n})$ directly affects the potential for panspermia, $x$ also has exceedingly strong influence on the prospects for panspermia. The range for $C_{n}$ extends from $\sim 10^{4-19}$ with the range of $x$ considered herein. 

Figure \ref{fig:f2} shows the effect of minimum shielding size $x$ on the ISM mass density $\rho_{m}(x)$. $\rho_{m}(x)$ decreases with $x$, the extent of which depends strongly on the number density slope $q$. The value of $\rho_{m}(x)$ for $q = 3.1$ overtakes both $q = 3.6$ and $q = 4.1$ at $x\sim 10$\,m. We see that the ISM mass density is relatively insensitive to the minimum shielding depth for $q=3.1$, whereas there is a strong dependence for $q=4.1$. Table \ref{tab:ism densities} lists values of $\rho_{m}(x)$ and $C_{m}$ given $q = 3.6$ and varying $x$. Note the negligible difference between the ejecta upper size bounds of $r = 1.5R_{\oplus}$ and $r = \infty$\, for $\rho_{m}(x)$ due to the relative scarcity of large ejecta.  Thus the upper limit could be approximated as $r = \infty$\ without significant impact on our model results.

\subsection{Consideration of 2I/Borisov} \label{sec2:Borisov}

We could make use of an occurrence rate number density for 2I/Borisov to constrain our model power-law exponent $q$. However, the number density $N_{o}$ for 2I/Borisov-like interstellar objects has not been estimated in the current literature. We can instead use our power-law model to invert this question and estimate a potential range for $N_{b}$, the number density of 2I/Borisov. We assume $3.1 < q < 4.1$ and $R_{b} \approx$ 0.35\,km, and solve for $N_{b}$ based on the previously established Equation \ref{Eq:1}. We find $\approx 2\times10^{-3} < N_{b} < 5\times10^{-4}$.

\section{Minimum Shielding Depth} \label{sec:shielding}

In Section \ref{subsec:3.1}, we derive an expression for the minimum ejecta shielding depth to protect extremophiles from supernova high-energy radiation based on the model from \citet{Sloan2017}. In Section \ref{subsec:3.2}, we apply the sphere packing method to estimate the average distance between an `Oumuamua-like ejecta and its nearest supernova progenitor. In Section \ref{subsec:3.3}, we compare the most probable ejecta composition for supporting the prospects of panspermia. In Section \ref{subsec:3.4}, we evaluate the minimum ejecta shielding depth and present additional results from this shielding depth model. 

\subsection{Minimum Shielding Depth Model} \label{subsec:3.1}

We assume that supernova high-energy gamma ray radiation are the dominant source of sterilization events for interstellar ejecta, as opposed to stellar fly-bys or diffuse background high-energy cosmic rays. Based on the model proposed by \citet{Sloan2017}, given an ejecta of size $r$\,[m] ($R_{e}$ in their work), minimum ejecta shielding depth $x$\,[m], and attenuation coefficient $\mu$\,[m$^{-1}]$, we have

\begin{equation} \label{Eq:13}
E_{f} = \frac{6000\pi R^{2}_{e}\left(e^{\mu x} - 1\right)}{\mu},
\end{equation}

where $E_{f}$\,[J] is the gamma ray energy that reaches the ejecta \citep{Sloan2017}. Extremophiles can generally survive radiation up to $E_{f} = 6000$\,J \citep{Sloan2017}. $E_{f}$ is dependent on the initial supernova emittance energy $E_{i}$\,[J]. On average, supernovas release $E_{i} = 10^{44}$\,J of energy \citep{1993A&A...270..223K, Bell2014}. $E_{f}$ is additionally dependent on the cross-sectional area of the ejecta $\sigma_{e}$\,[m$^{2}]$ and its distance to the nearest supernova $d$\,[pc]. Thus, we have

\begin{equation} \label{Eq:14}
E_{f} = E_{i}\left(\frac{\sigma_{e}}{4\pi d^{2}{\hat{d}}^{2}}\right)
\end{equation}

or

\begin{equation} \label{Eq:15}
E_{f} = E_{i}\left(\frac{\pi r^{2}}{4\pi d^{2}{\hat{d}}^{2}}\right).
\end{equation}

where ${\hat{d}} \approx 3.1 \times 10^{16}$\,[m/pc]. Substituting $E_{f}$ from Equation (\ref{Eq:15}) into Equation (\ref{Eq:13}) and solving for the minimum ejcta shielding depth, we have

\begin{equation} \label{Eq:16}
x \approx \frac{\ln{(E_{i}\mu) - \ln(24000\pi d^{2}{\hat{d}}^{2})}}{\mu}.
\end{equation}

\begin{deluxetable}{ccc}
\tablenum{2}
\tablecaption{Modeled ISM densities and ejecta collision statistics \label{tab:atten}}
\tablewidth{0pt}
\tablehead{
\colhead{Material} & \colhead{Attenuation Coeff. $\mu$} & \colhead{Min. Shielding $x$} \\
\colhead{Element} & \colhead{m$^{-1}$} & \colhead{m}
}
\decimalcolnumbers
\startdata
Silicate & 5.7 & 2.5\\
Nickel & 28.4 & 0.57 \\
Iron & 23.6 & 0.68  \\
Water-ice & 2.0 & 6.6  \\
\enddata
\tablecomments{This table shows the attenuation coefficient $\mu$\,[m$^{-1}]$ corresponding to silicate, nickel, iron, and water-ice. It also shows the minimum ejecta shielding depth $x$\,[m] for each material composition based on Equation (\ref{Eq:16}) \citep{NIST}.}
\end{deluxetable}
\vspace*{-\baselineskip} 

\subsection{Supernova Proximity} \label{subsec:3.2}

Next, we estimate the average distance between an ejecta and its nearest supernova progenitor. We are interested in ejecta originating from young, active star formation clusters because they are the densest stellar regions that contain the most supernova progenitors \citep{1996AJ....111.2017V, 10.1111/j.1365-2966.2004.07360.x}. Later in time, after star-forming regions and ejecta disperse to through the Milky-Way, the average distance to the nearest supernova progenitor is much larger, given a typical supernova rate in a galaxy on the order of once per century. We choose Orion A, a massive star-forming region, as our representative model of a typical dense star-forming region environment in our Galaxy.

First, we apply the sphere packing method to estimate the distance between adjacent stars in Orion A. The stellar density of Orion A $\approx$ $11$\,stars$\cdot$/pc$^{-3}$ given by the number of stars in Orion A divided by its volume, or $3000$\,stars$/270\,$pc$^{3}$, assuming a uniform density\citep{refId0}. Therefore, on average, there exists $\approx$ $0.09$\,pc$^{3}$ of non-overlapping volume between adjacent stars in Orion A, which we represent as spheres of radius $\sim$$0.3$\,pc. 

Next, the ratio of the number of stars with masses $<8 M_\odot$ that will not result in supernova explosions to the number of supernova progenitors with masses $>8 M_\odot$ in Orion A is governed by the initial mass function for star-forming region; this ratio in Orion A is $\approx 30:1$\citep{Chabrier_2003}.  Applying the sphere packing method, we estimate that the volume between adjacent supernova progenitors in Orion A as $30 \times 4/3 \pi (0.3)^{3}$\,pc$^{3}$. Setting this volume equal to that of a sphere and solving for its radius gives half the distance between adjacent supernova progenitors. Therefore, the distance between supernova progenitors in Orion A is $\sim0.93$\,pc $\times$ $2$ $\sim 2$\,pc. Again assuming these supernova progenitors are uniformly distributed in the star-forming region, and the interstellar ejecta are also distributed uniformly through the star-forming region, the ejecta would on average also be a distance $d\sim2$\,pc from the nearest supernova progenitor. We adopt 2 pc as our average supernova distance estimate, even though supernova progenitors can be centrally concentrated in multiple stellar systems in star forming regions and not uniformly distributed. For example, the supernova progenitors comprising the Trapezium in the Orion Nebula fit within a sphere with radius of $\sim0.25$\,pc within the larger nebula with a radius of $\sim$ 4\,pc. In this particular nebula of the Orion star-forming region, the average distance of ejecta in the nebula from the central cluster would be $\sim$2.75\,pc ($=3-0.25$\,pc), where $\sim$3\,pc corresponds to the radius for enclosing half the volume of the 4\,pc radius sphere.

\begin{figure}[ht!]
\plotone{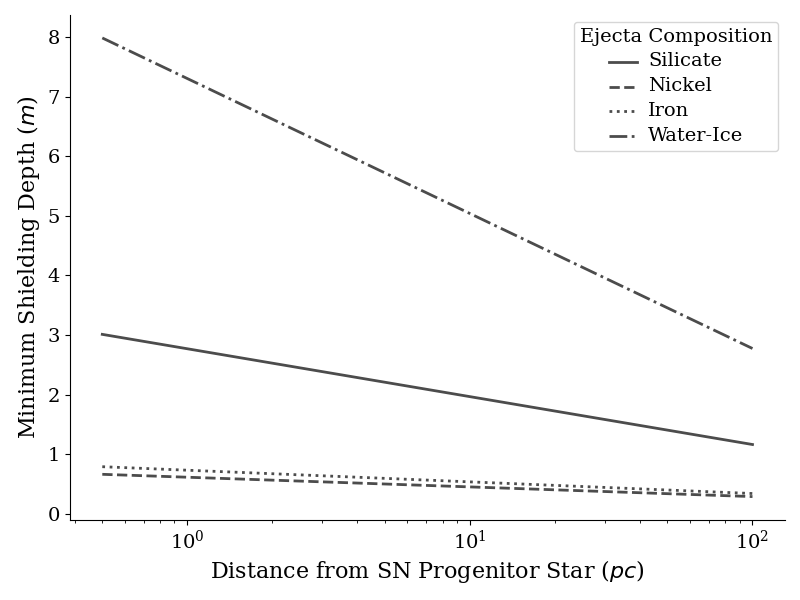}
\caption{Minimum ejecta shielding depth $x$\,[m] as it relates to the ejecta composition (silicate, nickel, iron, water-ice) and distance from the nearest supernova progenitor $d$\,[pc].\label{fig:f3}}
\end{figure}

\subsection{Ejecta Attenuation} \label{subsec:3.3}

Given the average distance from ejecta to supernova progenitors in a star forming region in the preceding section, we can soon calculate the expected average sterilization depths of ejecta given a choice of ejecta composition. The attenuation coefficient $\mu$\,[m$^{-1}$] of an ejecta is based on its composition. We consider the four most common elemental compositions of asteroids (chondritic, stony and metallic) and comets (water-ice) in our own Solar System: silicate, nickel, iron, and water-ice \citep{NASA}. Nickel has the highest attenuation coefficient of 28.4\,m$^{-1}$ and therefore requires the smallest minimum ejecta shielding depth $x$ to protect extremophiles, with water-ice providing the least attenuation \citep{NIST}. 

We next consider the relative number densities of interstellar ejecta of varying compositions. This is currently unknown and unconstrained -- the composition of `Oumuamua itself is highly debated with proposed exotic compositions that lack analogs in our Solar System such as molecular hydrogen ice and Nitrogen ice due to the lack of any observed tail of outgassing volatiles during its visit to our Solar System \citep{2019ApJ...872L..32M, 2020ApJ...900L..22L, 2020ApJ...896L...8S, 2021JGRE..12606706J}.  We make an assumption that the number density abundances and varying compositions of interstellar ejecta mirrors the content of minor bodies in our own Solar System. In the Solar System, the composition abundances are very dependent on formation location. The asteroid belt and inner Solar System minor bodies are thought to predominantly be of refractory heavier elements, with relatively higher condensation temperature compositions such as Nickel, Iron and Silicates, whereas the outer Kuiper Belt and Oort cloud are thought to be predominantly be of water-ice composition as a consequence of the relatively lower condensation temperature of water and other volatiles. However, recent exploration of asteroid belt bodies with missions such as Dawn and OSIRIS-Rex have shown significant water-ice and volatiles content, and indicate that in the past there has been significant orbital distance mixing of minor bodies in the Solar System from their original formation sites \citep{OSIRIS-Rex, OSIRIS-Rex-NASA, Padavic-Callaghan}.  In terms of number density abundances, the numbers of Oort Cloud and Kuiper belt objects could potentially outnumber asteroids by factors of $\sim$10$^{3-6}$ \citep[][and references therein]{2002aste.book..545D, 2012ApJ...761..150S,  2021Icar..35614256M, 2022arXiv220600010K, 2023AJ....166..242Z}.  Consequently, we conservatively assume an attenuation coefficient for gamma-ray radiation for water-ice on the assumption that water-ice ejecta dominate interstellar object number densities.

\subsection{Results: Minimum Shielding Depth} \label{subsec:3.4}

Applying $\mu = 2.0$\,m$^{-1}$ and $d \approx 2$\,pc to Equation (\ref{Eq:16}) yields a minimum ejecta shielding depth and size $x \sim 6.6$\,m. In Table \ref{tab:atten} we present a comparison of the attenuation coefficients $\mu$ and minimum ejecta shielding depths $x$ of silicate, nickel, iron, and water-ice. In Figure \ref{fig:f3} we plot the effect of an ejecta's distance from the nearest supernova $d$\,[pc] on the minimum ejecta shielding depth $x$. For Nickel and Iron, the minimum-shielding size has little dependence on distance to the nearest Supernova over the range shown of 0.5--100 pc, whereas for water-ice, the difference is only a factor of $\sim$2 in radius for the same distance range. Consequently our results for shielding depth are relatively insensitive to our choice of supernova proximity. Finally, in Table \ref{tab:ism densities} we present the ISM mass density $\rho_{m}(x)$, the total number of impact events $C_{n}(\Phi_{n})$, and the total collision mass $C_{m}(\Phi_{m})$ prior to the earliest fossilized evidence for life given minimum ejecta shielding depth and size $x\approx$ $6.6$\,m.  Here we see that the assumed collision rate is highly sensitive to the minimum shielding depth $x$ as a consequence of the power-law distribution of the number density of ejecta.

\section{Discussion} \label{sec:summary}

In Section \ref{subsec:4.1}, we place our findings in the context of the plausibility of panspermia. In Section \ref{subsec:4.2}, we compare our findings with that of other work. In Section \ref{subsec:4.3}, we present additional factors that may affect the plausibility of panspermia. 

\subsection{Implications on Panspermia} \label{subsec:4.1}

In this study, we assessed the mechanism for panspermia by estimating the number of impact events on Earth prior to the earliest fossilized evidence for life, that could have plausibly contained extremophiles that survived an interstellar transfer. Given minimum ejecta size $x \approx 6.6$\,m, we estimate $C_{n} \approx 1.9 \times 10^{5}$ for $q = 3.6$, and this should be regarded as an order-of-magnitude estimate. However, several other factors, most of which are unconstrained, must be considered to quantitatively gauge the plausibility of panspermia. 

First, consider $f_{B}$ as the probability that a randomly selected ejecta initially harbors microbial extremophile life. \citet{doi:10.1089/ast.2021.0187} hypothesizes that $10^{-13} \le f_{B} \le 10^{-10}$. Combining this estimate with the collision rate estimate from Table \ref{tab:ism densities}, the number of potential biologically active ejecta that impacted Earth prior to the earliest fossilized evidence for life is $\sim10^{-8} \le C_{n}f_{B} \le 10^{-5}$. The most conservative estimate where $q = 3.1$ gives $\sim10^{-9} \le C_{n}f_{B} \le 10^{-6}$, and the most liberal estimate where $q = 4.1$ gives $\sim10^{-8} \le C_{n}f_{B} \le 10^{-5}$. Overall, the number of biologically active ejecta that impacted Earth $\sim10^{-9} \le C_{n}f_{B} \le 10^{-5}$ over the range of values for $q$ considered herein.

Second, we must also consider $f_{su}$ and $f_{se}$, the fraction of ejecta that reach the surface of Earth upon impact and the fraction of extremophiles that successfully seed life, respectively, using the same convention as in \citet{doi:10.1089/ast.2021.0187}.We are explicitly computing the probability that viable organisms would be delivered to the surface of the Earth, and were able to then survive on the surface, and we denote this phenomenon as ``seeding" life as used in \citet{doi:10.1089/ast.2021.0187} These factors will lower $C_{n}f_{B}$ and reduce the likelihood of panspermia. Since we cannot constrain these factors to estimate a value for $C_{n}f_{B}f_{su}f_{se}$, we can instead place an upper-limit to the probability that panspermia seeded life on Earth is $C_{n}f_{B}f_{su}f_{se} \leq 10^{-5}$. Our upper limit for $C_{n}f_{B}f_{su}f_{se}$ is based on the assumption that both $f_{su} = 1$ and $f_{se} = 1$, which is under the most optimistic conditions that all microorganisms on an impact ejecta reach Earth’s surface ($f_{su} = 1$) and that ejecta successfully propagates life ($f_{se} = 1$). From our study, $10^{-9} \leq C_{n}f_{B} \leq 10^{-5}$. In this inequality, $10^{-5}$ is the maximum possible value for $C_{n}f_{B}$, which would be true if both $f_{su} = 1$ and $f_{se} = 1$ in our expression for $C_{n}f_{B}f_{su}f_{se}$.

If we instead assume more refractory materials interstellar ejecta, like nickel-iron metallic and silicate stony asteroid-like ejecta, we can lower the shielding depth required to $\sim$1 m, and increase $C_{n}$ by a factor of $\sim10^3$ to $C_{n}\approx2\times 10^{8}$, and thus $\sim10^{-6} \le C_{n}f_{B} \le 10^{-2}$ for varying values of q considered, corresponding to a non-negligible upper-limit of $C_{n}f_{B}f_{su}f_{se} \leq 10^{-2}$. After all, $\sim$95\% of meteors that survive impact on Earth and leave meteorite fragments are stony in composition, which would also seem to imply a higher $f_{su}$. However, this is misleading and incorrect. First, water-ice meteorite fragments would melt on a temperate Earth. Second, we would also have to reduce our number density constant $N_o$ for interstellar ejecta by a similar or larger factor of $\sim10^{3-6}$ to account for the relative number densities of interstellar ejecta as a function of composition under the assumption that it mirrors the relative number densities in our present Solar System - e.g. we are estimating stony interstellar ejecta to be $\sim10^{3-6}$ less common than water-ice ejecta.  Consequently, and coincidentally, we are left with an upper limit of $C_{n}f_{B}f_{su}f_{se} \leq 10^{-5}$ for refractory interstellar ejecta, the same as for water-ice ejecta.  

For the Earth in particular, we conclude that, independent of other hypotheses for the origins of life on Earth, panspermia remains improbable ($<0.001\%$) under the assumptions herein motivated by the increased rates of interstellar ejecta inferred from the discoveries of `Oumuamua and Comet Borisov. This result is robust against the particular choice of number density power-law $q$, ejecta composition, average supernova distance, and supernova sterilization depth. However, our computed probability for panspermia, when interpreted in the absolute, should be relative to the probabilities of other origin-of-life theories including prebiotic evolution and Mars transport. Therefore, we conclude that the true relative probability for panspermia remains unknown, but must be higher than our computed probability for panspermia on the basis that all other hypotheses for the origins of life have occurrence probabilities less than 1.

We do, however, find a surprising conclusion for panspermia for our Galaxy (and other galaxies) as a whole. There exists an estimated $\sim4\times10^9$ Earth-sized rocky planets in Habitable Zone (liquid surface water) orbits of Sun-like FGK stars, taken by multiplying $\eta_{\oplus}\sim0.1-0.25$ by the number of stars in our Galaxy \citep[][and references therein]{5ab6c203-4eeb-33f2-9854-ff00a02e6606, 2014ApJ...795...64F, 2015ApJ...809....8B, 2015ApJ...807...45D, 2019AJ....158..109H, 2021AJ....161...36B, 2021exbi.book....2G, 2023AJ....166..234B}. The fraction of these rocky planets that possess magnetic fields, atmospheres, and liquid surface water capable of supporting life is currently unconstrained and unknown, but our work implies as many as $10^4$ of these worlds in our Galaxy could be populated with life today via panspermia under the most optimistic assumptions that all of these worlds are capable of supporting ejecta-transported life, with Earth as one of the potential source planets. However, our model only accounts for the number of collisions in a 0.8 Gyr interval in the case of the Earth's existence. For Earth analogs in our Galaxy as a whole, some of them presumably as old as the Galaxy $\sim$13 Gyr, we pick up an additional factor of $\sim$10 increase in the number of collisions if we consider life-seeding over the history of the Galaxy, implying that as many as $10^5$ of these worlds in our Galaxy could be populated with life today via panspermia under the most optimistic assumptions. Herein we assumed a constant collision rate over Galactic time, which we revisit in ${\S}$\ref{subsec:4.3}.

\subsection{Comparison With Other Work} \label{subsec:4.2}

Following the discovery of `Oumuamua, \citet{siraj20222019} estimates the ISM ejecta number density $\rho_{n}$ for ejecta with sizes $> 0.5$\,m $\sim 10^{6}$. In contrast, based on Equation (\ref{Eq:2}), our model estimates $\rho_{n} \sim 10^{3-4}$. Additionally, \citet{Moro-Martín_2022} concludes that $\rho_{n}$ of ejecta with sizes $\sim 10^{-1}-10$\,m $\sim 10^{4}$ with $q \approx 3.6$, whereas our study concludes $\rho_{n} \sim 10^{3-4}$ also given $q \approx 3.6$. 

In addition, our analysis concludes that the maximum number of panspermia events on Earth prior to the earliest fossilized evidence for life $\sim10^{-8} \le C_{n}f_{B} \le 10^{-5}$, where $C_{n}f_{B}$ is the fraction of ejecta harboring extremophiles that collided with Earth. In comparison, \citet{doi:10.1089/ast.2021.0187} estimates the maximum number of panspermia events $\sim 10^{-6}$, an order of magnitude smaller than our upper-limit. Notably, our estimation for the maximum number of panspermia events has a range partly because we assume variable power law exponent $q$, where $3.1 < q < 4.1$, whereas \citet{doi:10.1089/ast.2021.0187} assumes a definitive power law exponent $q \approx 1.8$.

\subsection{Additional Factors for Panspermia} \label{subsec:4.3}

In this section, we discuss a number of additional factors that may impact the plausibility of panspermia that were not considered in our analysis. First, we consider the implications of the recently discovered rogue free-floating planets \citep{2023arXiv230308280S, Mróz2017}. Second, we address the impact that the Late Heavy Bombardment may have had on the ISM ejecta number density. Third, we consider the effects of the evolution of star formation rate as a function of time. Finally, we assess atmospheric grazing as a mechanism for panspermia.

First, the discovery of rogue-free floating planets (FFPs) suggests a significantly higher ISM ejecta number density than expected for large objects. \citet{Mróz2017} analyzed data from the OGLE-IV sky survey from 2010 to 2015 and found 6 short timescale events with $0.1 < t_{E} < 0.4$\,days, confirming the existence of terrestrial FFPs. The true number of FFPs is substantially larger than what is observed due to observational incompleteness \citep{2023arXiv230308279K}. The model proposed by \citet{2023arXiv230308280S} suggests that the number of FFPs is $\sim19$ times the number of planets in bound wide orbits, despite both regions having the same mass density.  This can represent a significant number of rogue worlds for extremophile transport panspermia, outnumbering the stars. However, we have not considered this size regime in our analysis, focusing instead on the small body ejecta. There is also no evidence that the power-law that governs the number density distribution for small bodies the size of `Oumuamua extends all the way to the regime of free-floating planets, and to the contrary a broken power-law is likely \citep[e.g., ][]{1984PASJ...36..357I, 2002aste.book..545D, 2021Icar..35614256M}.
 
Second, $\sim$4 Gyr ago, the Earth is thought to have experienced an unprecedented number of impact events that consequently ejected matter into the ISM \citep{Morbidelli_2007}, the era of Late Heavy Bombardment (LHB). \citet{1990LPICo.746....4B} estimates the number of impact events on Earth by finding the number of craters on the Moon. \citet{1990LPICo.746....4B} discovered that, if the LHB did occur, the rate of bombardments on Earth during the LHB was $\sim100-500$ greater than the present rate. Therefore, if similar LHB events occur in other planetary systems, the production of ISM ejecta number density could be substantially higher. Presumably, today's interstellar ejecta has reached a steady state throughout the Milky Way, but we speculate that at times in the past it could have been highly variable and stochastic given the proximity to other nearby, younger life-bearing stars.

Third, the star formation rate (SFR) may also alter the plausibility of panspermia. As more stars are formed, more mass will be ejected into the ISM in star formation regions, increasing the production of ISM ejecta number density \citep{Allen1993}. Similarly, as the low-mass stars like our Sun and less massive live many Gyr and are the most abundant spectral types, the total number of stars has increased with time, increasing the number of systems available to remove interstellar ejecta. Consequently, the net number of interstellar ejecta is a function of time in the difference between the ejecta production (source) and ejecta removal (sink) rates.  The production rate is proportional to the star-formation rate, and the removal rate is proportional to the number of stars, which is in turn proportional to the integral of the star-formation rate. Thus, we can introduce a time-dependent scaling factor to account for these time-dependent effects over the history of a Galaxy. We can express this as a non-linear factor $\delta = \alpha SFR(t) - \beta \int_{t_{i}}^{t_{f}} {SFR}dt$ that changes the ISM ejecta number density over time, where $\alpha$ and $\beta$ are constants. By measuring SFR rates in galaxies across a variety of redshifts and look-back times, various studies such as \citet{Yungelson_2000} have reconstructed the decreasing SFR with respect to time, while the number of stars is increasing with time. Therefore, both the production and removal mechanisms for interstellar ejecta imply that $\delta$ is decreasing with time and was higher in the past compared to the present. This in turn implies that the ISM ejecta number density was also higher in the past around the time of the Earth's formation, and we do not correct for that factor here in our analysis. 

Finally, atmospheric grazing occurs when a meteor grazes Earth's atmosphere and ``skips'' off back into space without colliding with the planet's solid surface. In the context of panspermia, an ejecta may seed life on Earth from atmospheric grazing as studied in \citet{siraj2020transfer}. According to \citet{siraj2020transfer}, when investigating the prospects for Venus--Earth panspermia, due to the immense friction and heat generated by atmospheric drag, extremophiles can only endure atmospheric altitudes $> 85$\,km for a 0.15 m radius asteroid (60 kg assuming 4g$\cdot$ cm$^{-3}$). In this study, however, we are considering much larger ejecta with embedded extremophiles in the core. We also do not account for atmospheric grazing because the collisional cross-section for a terrestrial planet atmosphere is $\sim2.5\times10^{-5}$ of the planet itself (0.5\% of the radius); direct collisions are far more common. Note, we ignore cases of planets with thick hydrogen-helium envelops that may nonetheless potentially also support life as has been recently suggested \citep{MolLous2022}.

\subsection{Potential Future Observational Constraints From Rubin} \label{subsec: 4.4}

Various studies predict that the highly-anticipated Rubin Observatory Large Synoptic Survey Telescope (LSST) will allow for the detection and detailed physical analysis (e.g. size) of substantially more interstellar objects \citep{10.1093/mnrasl/slad049, Jones2009, Mainzer_2015, Ivezić_2019, Rice_2019, Hoover_2022, Marčeta_2023}. 
\citet{Marčeta_2023}, for instance, simulated the detection of up to 70 interstellar objects with sizes 50-600 meters annually by the Vera Rubin Observatory assuming the same $0.1$\,au$^{-3}$ interstellar flux density for 'Oumuamua and Borisov-like objects. Additionally, \citet{10.1093/mnrasl/slad049} estimates with $90\%$ confidence that Rubin would detect another `Oumuamua-sized object within 5 years of its inception, and \citet{Rice_2019} concludes that Rubin may detect over 100 objects of size $> 1$\,m annually. If this does indeed come to fruition, Rubin could potentially enable us to further empirically constrain the flux density and power-law size distribution of ISM objects as explored in the existing literature and our model, at least in the diameter regime Rubin will be sensitive to \citep{Jones2009, Mainzer_2015, Vereš_2017, Ivezić_2019, Rice_2019, Hoover_2022, Marčeta_2023}. This will allow us to have more accurate estimations for the number density of ISM objects, the total number of impact events on Earth, and consequently, the probability that panspermia seeded life on Earth.

Additionally, such Rubin discoveries may enable the determination of the bulk compositional abundances \citep{Mainzer_2015, Ivezić_2019, Hoover_2022}, in particular, water-ice vs refractory vs more exotic materials. In this work, we assumed all water-ice, and determining the bulk compositional abundances of interstellar objects based on the discoveries of Rubin would allow for a more constrained model for the minimum size of interstellar objects, and thus, panspermia. Finally, as an exotic possibility, it may be possible from spectroscopy of a large enough sample of these ISM objects to look for any trace biomarkers or prerequisite organic chemical abundances for life (e.g. amino acids such as glycine) and the percentages of ISM objects containing these markers. This will enable us to replace some of the upper limits in our model with empirical measurements, as well as helping to understand the impacts of surface sterilization and chemical alteration from supernova flybys.

Overall, we present the discoveries of Rubin as an area of future work that would require completeness correction modeling as a function of object albedo (composition) and object diameter, and the survey design and sensitivity to recover the power-law distribution in this size regime.

\subsection{Panspermia Transfer via Interstellar Object Ejecta} \label{subsec: 4.5}

In this section, we discuss the implications for disseminating life via dust ejection from an interstellar object, with the dust remaining on the stellar system and eventually settling on a host planet. This discussion is motivated by the estimated mass-loss of Comet 2I/Borisov while it transited the inner Solar System.  \citet{Cremonese_2020} found a mass loss-rate of $Q_{d} = 35$\,kg/s from the tail of 2I/Borisov  when it was near perihelion.  It is conceivable that if some of this ejecta was life-bearing as it was released into our Solar System, some of that debris might eventually linger in our Solar System and intersect and settle into the upper Earth atmosphere.  We next evaluate how much mass might transfer to an Earth under this scenario.

The smallest grains ($<$ 0.3 microns approximately) will get blown out of the Solar System by radiation pressure \citep{ Chen_2001}. The remaining larger grains ejected from 2I/Borisov will be slowly dragged inward towards the Sun through Poynting-Robertson drag or stellar wind drag \citep{Plavchan_2005}. These dust grains have a lifetime of up to $10^{5}$ in our existing Solar System zodiacal dust cloud \citep{Chen_2001}. We note that originally the zodiacal dust cloud was thought primarily to originate from collisions of asteroids in our asteroid belt, but recently \citet{https://doi.org/10.1029/2020JE006509} has demonstrated that a significant fraction of this dust may in fact be ejected by Mars during planet-wide dust storms (as an aside, an interesting related prospect for panspermia life transfer within the Solar System).

We next estimate the total time that 2I/Borisov spends within a 4\,AU radius from the Sun, which equates to $d/v_{b}$ = ($2.1\times2\,$AU)/($32.2$ km/s) $\approx 1.95\times 10^{7}$ seconds $ \approx 256$ days, where 2.1\,AU $= 4\,$AU - $1.9\,$AU, and $1.9\,$AU is the perihelion distance for 2I/Borisov, and $v_{b}$ is the velocity of 2I/Borisov \citep{Guzik2020}. Thus, multiplying the dust production rate by time duration, we have the total mass of the dust grains produced by 2I/Borisov interior to 4\,AU of $ M_{d} \approx 6.8\times10^{8}$ kg.

We next assume this amount of dust is deposited in a circumstellar ring at 4\,AU of random inclination (not necessarily aligned with the ecliptic). The amount of dust that could eventually reach the Earth can be approximated by taking the ratio between the cross-sectional area of the Earth and a spherical shell at 1\,AU, $M_{f} = M_{d}(R_{\oplus}^{2}/1\,AU^{2})h_{\oplus}h_{\odot}$, where $h_{\oplus}$ and $h_{\odot}$ is the gravitational focusing contributions of the Earth and the Sun, respectively. Hence, we have $M_{f} \approx 5$\,kg.  This is the potential interstellar dust mass deposited for every interstellar object of this size similar to 2I/Borisov that comes into the inner Solar System. Using our model, this would correspond to a total of mass of 5.4$\times10^{9}$\,kg in 800 Myr.

The next consideration is whether or not this dust can be life-bearing, since it comes from the surface of the interstellar object, which we assume is sterilized down to some depth of several meters. To confirm how much of the surface of an interstellar object erodes from this mass loss, we approximate the thickness of the spherical shell where the dust material ejected in 2I/Borisov by solving for $R_{b}-R{s}$ from the following equation ($R_{b}$ is the spherical radius of 2I/Borisov and $R_{s}$ is the radius of 2I/Borisov after mass loss following the ejection of materials), $\frac{R_{s}^{3}}{R_{b}^3} = \frac{M_{b} - M_{f}}{M_{b}}$. We first evaluate $M_{b}$, the mass of 2I/Borisov, to be $4/3\pi(R_{b})^{3}\rho = 4/3\pi(350\,$m$)^{3}500$ kg/m$^{3} \approx 9\times 10^{10}\,$kg \citep{Jewitt_2020}. Hence, we get $R_{s} = 349.111\,$m, and therefore, $R_{b} - R_{s} \approx 1\,$m. Since the thickness of the ejected layer, 1\,m, is less than the minimum supernova gamma radiation shielding depth of 6.6\,m, based on our model, we conclude that the ejected dust grains would most likely be sterilized. One caveat, however, would be if the life in the interior of the comet is able to “recolonize” the exterior of the comet after sterilization event(s). Herein we assume that a sterilization event leaves the surface sterilized permanently; we don’t consider the scenario of life repopulating the outer layers of the comet.

We conclude that the mass transfer via dust grains ejected from interstellar objects is large relative to the mass from direct collisions. However, it would appear that much of this dust would be from the sterile surfaces of the interstellar objects, unless any life-bearing interstellar objects are able to repopulate the surface after sterilization event(s).

\section{Conclusions \& Future Work} \label{sec:concf}

In Section \ref{sec:conclusions}, we summarize our study's methodology, findings, and discussion about the implications of `Oumuamua on panspermia. In Section \ref{Future Work}, we propose potential areas for future work to build on.

\subsection{Conclusions} \label{sec:conclusions}
In this study, given the recent discovery of `Oumuamua, we revisited the plausibility of panspermia. First, we modeled the ISM ejecta number density and the ISM ejecta flux rate with `Oumuamua as an anchor/normalization. Next, we modeled the ISM ejecta mass density and the ISM mass flux. We then evaluated the minimum radius of ejecta required to shield extremophiles from supernova gamma-ray radiation and achieved a result of $x \approx 6.6$\,m. With that size constraint, we calculated the total number of collisions on Earth after its formation and prior to emergence of life, $C_{n} \approx 2 \times 10^{5}$. Next, we considered poorly constrained factors such as the fraction of ejecta harboring extremophiles, $f_{B}$, the fraction of ejecta that reach Earth's surface, $f_{su}$, and the chance that the extremophiles seed life, $f_{se}$ \citep{doi:10.1089/ast.2021.0187}. We formulate that the chance that panspermia seeded life on Earth is governed by the product $C_{n}f_{B}f_{su}f_{se}$. Since $f_{su}$ and $f_{se}$ are unknown, we evaluate the maximum chance that panspermia seeded life on Earth in particular to $C_{n}f_{B}f_{su}f_{se} \le 10^{-5}$. However, when interpreted in the absolute, the probability for panspermia should be relative to other origin-of-life theories, therefore the true probability for panspermia remains unknown and is higher than what we computed. We do, however, find that, applying our estimation of the probability for panspermia, given the preponderance of Earth-sized worlds in Habitable Zone orbits of Sun-like stars, panspermia remains a plausible mechanism for seeding life for up to $10^{5}$ Earth-analogs in our Galaxy under optimistic assumptions. We compared our results with other studies and address similarities and differences. Finally, we introduced and discussed several additional potential factors affecting the prospects panspermia not accounted for in our model, including the implications of rogue free-floating planets, the Late Heavy Bombardment, the historical evolution of the star formation rate, findings from the Vera Rubin Observatory, and the prospects for the transfer of life via dust-grained sized ejecta from an interstellar object during a flyby.

\subsection{Future Work} \label{Future Work}
To build on this study, future work can revise the ISM ejecta number density and mass density models by analyzing and incorporating data from free-floating planets. In addition, future work can estimate the extent to which the ISM ejecta number density increases if other planetary systems experience periods of heavy impact events akin to the Late Heavy Bombardment. We also propose that future work can adjust the ISM ejecta number density according to the decreasing star formation rate as a function of Galactic time. Finally, future work can constrain the power law size distribution, abundance, and composition of interstellar ejecta based on findings from the Vera Rubin Observatory and 2I/Borisov.

\begin{acknowledgements}
The authors thank the Aspiring Summer Scientists' Internship Program (ASSIP) for providing the opportunity for high-school intern David Cao to work alongside Professor Summers and Professor Plavchan in the George Mason University Physics \& Astronomy department. We also acknowledge the use of Matplotlib from \citet{Hunter:2007} to create all figures displayed in this study. PPP would like to acknowledge support from NASA (Exoplanet Research Program Award \#80NSSC20K0251, TESS Cycle 3 Guest Investigator Program Award \#80NSSC21K0349, JPL Research and Technology Development, and Keck Observatory Data Analysis) and the NSF (Astronomy and Astrophysics Grants \#1716202 and 2006517), and the Mt Cuba Astronomical Foundation. 
The authors thank Adam Hibberd for the suggestion of adding the Sun's gravitational focusing effect.
\end{acknowledgements}

%



\bibliography{ms}{}
\bibliographystyle{aasjournal}



\end{document}